\documentclass[structabstract]{aa}  
\usepackage{graphicx}
\usepackage{txfonts}
\usepackage{natbib}
\usepackage{verbatim}
\usepackage{comment}
\usepackage{color}
\usepackage{bm}
\newcommand{\so}{$\sigma$~Orionis~}
\newcommand{\metcol}{$H-CH_{\rm 4on}$~}
\newcommand{\met}{$CH_{\rm 4on}$~}
\newcommand{\cero}{S\,Ori\,70~}
\newcommand{\tres}{S\,Ori\,73~}
\newcommand{\master}{J0538$-$0213~}
\newcommand{\mj}{\,M$_{\rm Jup}$~}
\bibpunct{(}{)}{;}{a}{}{,} 

\begin{document}
\title{Characterization of the known T-type dwarfs towards the $\sigma$ Orionis cluster}
\subtitle{}
\author{K. Pe\~na  
Ram\'irez\inst{1,2} \and M.R. Zapatero Osorio\inst{3} \and V.J.S. B\'ejar\inst{4,5}}
\institute{Instituto de Astrof\'isica. Pontificia Universidad Cat\'olica de Chile (IA-PUC), E-7820436 Santiago, Chile. \\ \email{kpena@astro.puc.cl} \and Millennium Institute of Astrophysics, Santiago, Chile \and Centro de Astrobiolog\'ia (CSIC-INTA), Carretera de Ajalvir km 4, E-28850 Torrej\'on de Ardoz, Madrid, Spain. \and Instituto de Astrof\'isica de Canarias (IAC), C/$.$ V\'ia L\'actea s/n, E-38205 La Laguna, Tenerife, Spain. \and Universidad de La Laguna, Tenerife, Spain.\\}
\date{Received 18 August 2014 / Accepted 12 November 2014}

\abstract
{}
{ The detailed study of T-type candidate members of the young $\sigma$\,Orionis cluster ($\sim$3\,Myr, $\sim$352\,pc, solar metallicity) is fundamental to properly assess the objects' cluster membership and their contribution to the definition of the substellar mass function.}
{A total of three T-type candidates (S\,Ori\,70, S\,Ori\,73, and S\,Ori\,J053804.65$-$021352.5) lying in the line of sight towards $\sigma$\,Orionis were characterized by means of near-infrared photometric, astrometric, and spectroscopic studies. $H$-band methane images were collected for all three sources 
and an additional sample of 15 field T-type dwarfs using the LIRIS instrument on the 4.2\,m William Herschel Telescope (WHT). $J$-band spectra of resolution of $\sim$500 were obtained for S\,Ori\,J053804.65$-$021352.5 with the ISAAC spectrograph on the 8\,m Very Large Telescope (VLT), and $JH$ spectra of resolution of $\sim$50 acquired with the Wide Field Camera 3 (WFC3) on board the Hubble Space Telescope (HST) were employed for the spectroscopic classification of S\,Ori\,70 and 73. Accurate proper motions with a typical uncertainty of $\pm$3 mas\,yr$^{-1}$ and a time interval of $\sim$7--9\,yr were derived using old images and new data collected with ISAAC/VLT and WFC3/HST. }
{Using\textbf{ }the LIRIS observations of the field T dwarfs, we calibrated this imager for T spectral typing via methane photometry. The three S\,Ori objects were spectroscopically classified as T4.5\,$\pm$\,0.5 (S\,Ori\,73), T5\,$\pm$\,0.5 (S\,Ori\,J053804.65$-$021352.5), and T7$^{+0.5}_{-1.0}$ (S\,Ori\,70). These spectral types agree with the measured $H$-band methane colors. The similarity between the observed $JH$ spectra and the methane colors and the data of field ultra-cool dwarfs of related classifications suggests that S\,Ori\,70, 73, and S\,Ori\,J053804.65$-$021352.5 do not deviate significantly in surface gravity in relation to the field. Additionally, the detection of K\,{\sc i} at $\sim$1.25\,$\mu$m in S\,Ori\,J053804.65$-$021352.5 points to a high-gravity atmosphere. Only the $K$-band reddish nature of \cero may be consistent with a low-gravity atmosphere. The proper motions of S\,Ori\,70 and 73 are measurable and are larger than that of the cluster by $>$3.5-$\sigma$. The proper motion of S\,Ori\,J053804.65$-$021352.5 is consistent with a null displacement. These observations suggest that none of the three T dwarfs is a likely $\sigma$\,Orionis member, and that either planetary-mass objects with masses below $\sim$4\,M$_{\rm Jup}$ may not exist free-floating in the cluster or they may lie at fainter near-infrared magnitudes than those of the targets (i.e., $H > 20.6$\,mag), thus remaining unidentified to date. We determined the volume density of field T4--T7 dwarfs to be $\ge$2.8\,$\pm$\,1.6\,$\times$\,10$^{-3}$ pc$^{-3}$ from a survey that covered 2798.4 arcmin$^2$ and was complete up to a distance of 119\,pc.}
{}
\maketitle

\section{Introduction}
The determination of the minimum mass limit for the collapse and fragmentation of clouds is crucial for testing the different formation models that try to explain the existence of brown dwarfs and planetary mass objects. This minimum mass value remains poorly unconstrained by the theory and is observationally undetected. The different formation scenarios locate it in the range 1--7\mj \citep{low76,whitworth06,stamatellos07,boss11}. Considering
the comparison with field dwarf calibrators, along with the predictions made by evolutionary models \citep{baraffe03}, it may be estimated that low mass values, such as the ones considered, correspond to sources with temperatures of $\leq$1500\,K in regions with less than 10\,Myr. These physical parameters overlap with the methane T-type regime of very young star forming regions.

Several T-type objects have been proposed as members of very young star clusters and associations ($<$15\,Myr): IC\,348 \citep{burgess09}, $\rho$ Ophiuchus \citep{haisch10,marsh10,geers11}, Upper Scorpius \citep{lodieu11usco}, and Serpens \citep{spezzi12}, although the membership of these candidates is still under debate \citep{alvesdeoliveira10,lodieu13}. With an age of 100\,$\pm$\,30\,Myr, the likely T-type AB\,Doradus members CFBDSIR\,J214947.2$-$040308.9 \citep{delorme12} and the companion GU\,Psc\,b \citep{naud14} stand out. In the \so cluster --- with an age of around 3\,Myr \citep{zapatero02age,sherry08}, low extinction \citep{lee68,bejar04a}, solar metallicity \citep{hernandez08}, and distance of $\sim$300--450\,pc \citep{brown94,perryman97,sherry08,caballero08distance} --- the characterization of the known T-type dwarfs lying in the cluster direction is important in order to determine the planetary-mass function and their impact on a possible turnover (or a mass cut--off) around 4\mj \citep{bihain09,penaramirez12}. 

Here, we assessed the \so membership of three T-type candidates known towards the direction of the cluster: \cero \citep{zapatero02sori70}, \tres \citep{bihain09}, and S\,Ori\,J053804.65$-$021352.5 \citep[hereafter J0538$-$0213]{penaramirez12}. We presented the first combined astrometric and spectroscopic study of J0358$-$0213. For all three T-type objects together with a sample of field dwarfs ranging from T0 through T7, we collected methane images from which we inferred $H-CH_{\rm 4on}$ colors. We provided a calibration of the $H$-band methane colors against spectral subtypes. For \cero and S\,Ori\,73, we revised their proper motions and showed their low-resolution near-infrared spectra. In Sect.~\ref{targets},  we summarized the current knowledge of our targets prior to this work. In Sects.~\ref{data} and \ref{anal}, the observations and the main results are described. A discussion and final remarks are given in Sect.~\ref{discussion}.


\section{Target selection \label{targets}}
\subsection{T-type sources towards the \so cluster}
Three\footnote{Besides the three T-type S\,Ori targets in the present work, S\,Ori\,69 \citep{zapatero00} was initially typed as a T0 dwarf based on poor signal--to--noise ratio near-infrared spectra \citep{martin01}. Despite the depth of their explorations, neither \citet{caballero07} nor \citet{penaramirez12} detected S\,Ori\,69, which suggests that this object is fainter than was indicated in the discovery paper. Since the real nature of this source is still unclear, we consider S\,Ori\,70, S\,Ori\,73, and \master as the T-type sources in the direction of the \so cluster.} T-type \so  candidate members were identified:
 
\begin{itemize}
 \item S\,Ori\,70: Photometrically discovered by \citet{zapatero02sori70} in an area of 55.4\,arcmin$^2$. These authors classified \cero as a T5.5\,$\pm$\,1 type dwarf and suggested that it may be an isolated planetary mass candidate from its $HK$ low-resolution spectrum. Burgasser et al$.$ (2004) argued that this object is probably a foreground field T6--T7 dwarf based on $J$-band near-infrared spectroscopy. Its astrometric and photometric properties were discussed by \citet{zapatero08}, \citet{scholz08}, and \citet{penaramirez11}. S\,Ori\,70 displays red $J-K$ and \textit{Spitzer} colors qualitatively in agreement with the theoretical predictions for solar-metallicity, low-gravity atmospheres, and/or the possible presence of a surrounding disk \citep{zapatero08, luhman08, scholz08}. These aspects would support the young age of S\,Ori\,70. The proper motion determined by \citet{penaramirez11} lies at $\sim$4.5-$\sigma$ from the motion of the central multiple star $\sigma$ Ori measured by \textit{Hipparcos} \citep{perryman97}.
  
 \item S\,Ori\,73: Photometrically discovered by \citet{bihain09} in an area of 840\,arcmin$^2$, \tres is $\sim$0.5 mag fainter than \cero in the $H$ band. The presence of methane absorption in its atmosphere was confirmed with deep methane-filter images obtained by \citet{penaramirez11}. These authors estimated a spectral type of T4 based on the methane color of S\,Ori\,73. Spectroscopic data were presented in a conference paper by \citet{lucas13}. The near- and mid-infrared colors and the proper motion of \tres (which deviates by $\sim$7.4-$\sigma$ from the \textit{Hipparcos} proper motion of the $\sigma$ Ori central star) suggest that it has a high-gravity atmosphere similar to field dwarfs of related spectral classification.
  
 \item J0538$-$0213: Photometrically identified in the 2798.4-arcmin$^2$, multiwavelength survey of \citet{penaramirez12}, this object has colors (from the $Z$ band through 3.6\,$\mu$m) fully compatible with an early- to mid-T-type dwarf. It is $\sim$0.8 mag brighter than \cero in the $H$ band. No astrometric or spectroscopic data of \master are available in the literature.
 
 \end{itemize}

\subsection{Field T dwarfs}
In addition to the aforementioned targets, we also included a sample of 15 known field T-type dwarfs that are clearly unrelated to the \so cluster in our methane imaging campaigns. All are detected in the Two Micron All Sky Survey (2MASS; \citealt{skrutskie06}) and the UKIRT Infrared Deep Sky Survey (UKIDSS; \citealt{lawrence07}) as part of the Large Area Survey's T dwarf program. They have spectral types in the interval T0--T7 measured from near-infrared spectra and will be used to calibrate the observed \metcol color as a function of spectral type. Table~\ref{liris} provides their names, spectral types, and discovery papers.

\section{Observations}\label{data}
\subsection{Near--infrared spectroscopy}\label{spectroscopy}
$J$-band near-infrared spectroscopy of J0538$-$0213 was obtained using the Infrared Spectrometer And Array Camera (ISAAC; \citealt{moorwood98}) installed on the Nasmyth A focus of the third telescope of the Very Large Telescope (VLT) array sited on Cerro Paranal (Chile). ISAAC is an imager and spectrograph that covers the wavelength interval 1--5\,$\mu$m. For our observations we used the short wavelength arm that is equipped with a 1024\,$\times$\,1024 Hawaii Rockwell detector with a pixel size of 0\farcs147 and covering 1--2.5\,$\mu$m. Our data were obtained in the low-resolution mode with a slit width of 1\farcs0 and centered at 1.25\,$\mu$m. This instrumental configuration yielded a spectral nominal dispersion of 3.49\,\AA\,pix$^{-1}$, a resolving power of about 500 at the central frequency, and a wavelength coverage of 1.09--1.42\,$\mu$m. Because of the faint nature of our targets and because Earth atmospheric water vapor absorption is very strong redward of 1.34\,$\mu$m, the useful wavelength coverage is 1.11--1.34\,$\mu$m. ISAAC observations of J0538$-$0213 were collected with a seeing of 0\farcs7--1\farcs0 on 2012 December 4.

J0538$-$0213 and a bright reference star ($J$=17.5\,mag) at a separation of 17\farcs1 were acquired through the $J$-band filter and simultaneously aligned on the 120\farcs0-length slit. Individual 600\,s exposures were obtained with the source at two nod positions separated typically by 10\farcs0. The target was observed in an ABBA nodding pattern twice, yielding a total on-source integration time of 1.33\,h. To account for absorption by the Earth's atmosphere, a telluric standard of spectral type B was observed immediately after the target and as close to the same air mass as possible, within $\pm$0.1 air masses. Reduction of the raw data was accomplished using {\sc iraf}\footnote{The Image Reduction and Analysis Facility ({\sc iraf}) is distributed by National Optical Astronomy Observatories, which is operated by the Association of Universities for research in Astronomy, Inc., under contract to the National Science Foundation.}. Pairs of nodded target frames were subtracted to remove the background emission contribution and then divided by the corresponding flat field. Individual frames were registered using the bright reference star and stacked together to produce higher signal-to-noise spectra. Spectra were optimally extracted and wavelength calibrated using the terrestrial sky emission lines to a precision of about 10--20\%~of the nominal dispersion. After removal of the intrinsic features (typically hydrogen lines) of the B-type star, the calibration spectra were divided into the corresponding target spectra to remove telluric absorptions and instrumental spectral response. Finally, the data were multiplied by a black body curve of 32000\,K to restore the spectral slope. 

S\,Ori\,70 and 73 have publicly available spectroscopic data from the Mikulski Archive for Space Telescopes. As part of program number 12217 (principal investigator: P$.$ Lucas, see Lucas et al$.$ 2013), low-resolution, slitless spectra were obtained using the G141 grism and the Wide Field Camera 3 (WFC3) on board the Hubble Space Telescope (HST) on 2010 Sep 5 (S\,Ori\,70) and Oct 6 (S\,Ori\,73). The integration time was 40\,min for each target, which corresponded to the duration of one orbit. Observations followed a four-point dither pattern to remove cosmic rays. Data were reduced using the aXe\footnote{http://axe-info.stsci.edu} software package \citep{kummel09}, which included flat-field correction, background subtraction, optimal extraction of the spectra using an aperture of 6\,pix (or 0\farcs8, thus ensuring that about 90\%~of the flux is recovered\footnote{http://www.stsci.edu/hst/wfc3/documents/WFC3\_aXe\_cookbook.pdf, version of 2014 August}), and wavelength and flux calibrations. The final WFC3 $JH$ spectra have a nominal dispersion of 46\,\AA\,pix$^{-1}$, a resolving power of about 50 at the central frequency (1.39\,$\mu$m), and a useful wavelength coverage of 1.08--1.70\,$\mu$m.


\begin{table}
\caption{Journal of LIRIS imaging observations. \label{liris}}
\centering
\scriptsize
\tabcolsep=0.09cm
\begin{tabular}{llllccl}
\hline\hline
\noalign{\smallskip}
Object & SpT\tablefootmark{a} & Filter &  Date & Exp$.$ time & Seeing & Ref. \\
& & & & (s) & (\arcsec) & \\
\hline
\noalign{\smallskip}
S\,Ori\,70\tablefootmark{b}            & T5.5  & $H$            & 2009 Dec 15 & 9$\times$28$\times$10 & 1.57  & 11 \\
                                               &       & $CH_{\rm 4on}$  & 2009 Dec 15 & 9$\times$16$\times$20 & 0.90  &   \\
S\,Ori\,73                                     &       & $CH_{\rm 4on}$  & 2009 Jan 14 & 9$\times$21$\times$20 & 0.87  & 12  \\ 
S\,Ori\,J053804.65$-$021352.5\tablefootmark{b} &       & $H$            & 2011 Dec 31 & 9$\times$12$\times$15 & 1.50  & 13  \\ 
                                               &       & $CH_{\rm 4on}$  & 2011 Dec 31 & 9$\times$36$\times$15 & 1.13  &   \\
\hline
\noalign{\smallskip}
2MASS\,J072718.24$+$171001.2                   & T7.0  & $H$            & 2011 Dec 30 & 9$\times$8$\times$10  & 1.20  &  3     \\
                                               &       & $CH_{\rm 4on}$  & 2011 Dec 30 & 9$\times$8$\times$10  & 1.20  &          \\ 
2MASS\,J075547.95$+$221216.9                   & T5.0  & $H$            & 2011 Dec 30 & 9$\times$8$\times$10  & 1.23  &  3      \\
                                               &       & $CH_{\rm 4on}$  & 2011 Dec 30 & 9$\times$8$\times$10  & 1.53  &         \\
2MASS\,J075840.37$+$324724.5                   & T2.0  & $H$            & 2011 Dec 30 & 9$\times$8$\times$10  & 1.30  &  4     \\
                                               &       & $CH_{\rm 4on}$  & 2011 Dec 30 & 9$\times$8$\times$10  & 1.50  &          \\
2MASS\,J104753.85$+$212423.4                   & T6.5  & $H$            & 2011 Dec 31 & 9$\times$6$\times$15  & 1.25  &  1      \\
                                               &       & $CH_{\rm 4on}$  & 2011 Dec 31 & 9$\times$12$\times$15 & 1.00  &          \\
2MASS\,J110611.97$+$275422.5                   & T2.5  & $H$            & 2011 Dec 30 & 9$\times$8$\times$10  & 1.00  &  7      \\
                                               &       & $CH_{\rm 4on}$  & 2011 Dec 30 & 9$\times$8$\times$10  & 0.90  &         \\
2MASS\,J120747.17$+$024424.9                   & T0.0  & $H$            & 2011 Dec 30 & 9$\times$16$\times$10 & 1.80  &  2,4    \\
                                               &       & $CH_{\rm 4on}$  & 2011 Dec 30 & 9$\times$8$\times$10  & 1.83  &         \\
2MASS\,J152039.74$+$354621.0                   & T0.0  & $CH_{\rm 4on}$  & 2011 Dec 31 & 9$\times$12$\times$10 & 0.70  &  6      \\
2MASS\,J154614.61$+$493211.4                   & T2.5  & $H$            & 2011 Dec 31 & 9$\times$6$\times$10  & 0.70  &  8     \\
                                               &       & $CH_{\rm 4on}$  & 2011 Dec 31 & 9$\times$6$\times$10  & 0.65  &         \\
2MASS\,J000013.54$+$255418.0                   & T4.5  & $H$            & 2007 Dec 14 & 9$\times$1$\times$4   & 0.70  &  4     \\
                                               &       & $CH_{\rm 4on}$  & 2007 Dec 14 & 9$\times$1$\times$10  & 0.75  &         \\
2MASS\,J003451.57$+$052305.0\tablefootmark{c}  & T7.0  & $H$            & 2009 Jan 14 & 9$\times$6$\times$10  & 0.75  &  5     \\
                                               &       & $CH_{\rm 4on}$  & 2009 Jan 14 & 9$\times$3$\times$20  & 0.75  &         \\
ULAS\,J081948.10$+$073323.2\tablefootmark{c}   & T6.0p & $H$            & 2007 Dec 16 & 9$\times$14$\times$4  & 1.00  &  10    \\
                                               &       & $CH_{\rm 4on}$  & 2007 Dec 16 & 9$\times$5$\times$10  & 1.12  &         \\
ULAS\,J085139.00$+$005341.0\tablefootmark{c}   & T4.0  & $H$            & 2007 Dec 16 & 9$\times$14$\times$4  & 0.87  &  10    \\
                                               &       & $CH_{\rm 4on}$  & 2007 Dec 16 & 9$\times$5$\times$10  & 0.85  &         \\
ULAS\,J085342.94$+$000651.8\tablefootmark{c}   & T6.0p & $H$            & 2007 Dec 16 & 9$\times$7$\times$4   & 0.65  &  10    \\
                                               &       & $CH_{\rm 4on}$  & 2007 Dec 16 & 9$\times$5$\times$10  & 0.65  &          \\
ULAS\,J085910.69$+$101017.1\tablefootmark{c}   & T7.0  & $H$            & 2007 Dec 16 & 9$\times$14$\times$4  & 1.25  &  9     \\
                                               &       & $CH_{\rm 4on}$  & 2007 Dec 16 & 9$\times$5$\times$10  & 1.25  &         \\
ULAS\,J094516.40$+$075545.6\tablefootmark{c}   & T5.0  & $H$            & 2007 Dec 16 & 9$\times$9$\times$5   & 0.80  &  10    \\
                                               &       & $CH_{\rm 4on}$  & 2007 Dec 16 & 9$\times$4$\times$10  & 0.87  &         \\
\hline
\end{tabular}
\tablefoot{ 
The top panel lists the T-type objects in the direction of the \so cluster. The bottom panel contains 15 field T-type dwarfs.
\tablefoottext{a}{Spectroscopic spectral types taken from the literature.}
\tablefoottext{b}{$H$-band photometric calibration was performed using the VISTA Orion data presented in \citet{penaramirez12}.}
\tablefoottext{c}{$H$-band photometric calibration was performed using the DR8 UKIDSS catalog. For all other sources, 2MASS data were employed.}
}
\tablebib{(1) \citet{burgasser99}; (2) \citet{hawley02}; (3) \citet{burgasser02}; (4) \citet{knapp04}; (5) \citet{burgasser04}; (6) \citet{chiu06}; (7) \citet{looper07}; (8) \citet{metchev08}; (9) \citet{pinfield08}; (10) \citet{burningham10}; (11) \citet{zapatero02sori70}; (12) \citet{bihain09}; (13) \citet{penaramirez12}.
}
\end{table}

\subsection{Near--infrared imaging}\label{methane_imaging}
We obtained $H$-band and methane photometric data using the Long-slit Intermediate Resolution Infrared Spectrograph (LIRIS; \citealt{manchado04}) on the William Herschel Telescope located at the Observatorio Roque de los Muchachos in La Palma, Canary Islands, Spain. This camera has a HAWAII detector of 1024\,$\times$\,1024 pix with a plate scale of 0$\farcs$25 projected onto the sky. In imaging mode, LIRIS has a field of view of 4.27\,$\times$\,4.27\,arcmin$^2$. The set of filters includes a narrowband filter named \met with a passband of 1.64--1.74\,$\mu$m. The LIRIS \met filter is located at the red end of the $H$ band and within the methane absorption feature observed in T-type dwarfs. Typical seeing conditions during the LIRIS observations ranged from 0\farcs65 to 1\farcs83 and the weather was clear. 

We collected individual LIRIS frames of the S\,Ori objects and the 15 field T0--T7 dwarfs (see Section 2.2) using the $H$ and \met filters and following a nine-point dithering pattern with typical offsets of 15\farcs0 in right ascension and declination. Table~\ref{liris} shows the journal of the LIRIS observations, which includes target names, observing dates, seeing (as measured from the final reduced images), and exposure times (number of dithers $\times$ number of readouts per dither position $\times$ individual exposure). Raw data were corrected for flat field cosmetics, which incorporated a correction for a gradient effect using the task \textit{licvgrad} within the LIRIS data reduction package \textit{lirisdr}\footnote{www.iac.es/galeria/jap/lirisdr/LIRIS\_DATA\_REDUCTION.html as of 2014 August}. Individual frames were aligned and stacked to produce deep $H$- and $CH_{\rm 4on}$-band images.

Aperture and point spread function (PSF) photometry was obtained for all of our S\,Ori and field targets plus an additional $>$15 isolated, unresolved objects within the field of view of all targets. These additional sources were used for photometric and astrometric calibration purposes, as explained below. We used the {\sc iraf daophot} package for performing the photometric measurements. $H$-band instrumental magnitudes were converted into observed magnitudes using objects from 2MASS and UKIDSS (eighth data release, DR8) in common with our data. UKIDSS uses the UKIRT Wide Field Camera (WFCAM; \citealt{casali07}) and a photometric system described in \cite{hewett06}. The 2MASS and UKIDSS catalog photometry was converted to the Mauna Kea Observatory photometric system using the expressions given in \citet{leggett06}. All sources employed for the photometric calibration of our LIRIS $H$-band data have a precision better than $\pm$0.1~mag. The typical error of the $H$-filter photometric calibration is $\pm$0.04\,mag. The LIRIS \met filter was calibrated relative to the $H$ band following the procedure described in \citet{goldman10} and \citet{penaramirez11}. Objects within the field of view of our targets displaying $H$\,=\,13--16\,mag (probably Galactic stars of spectral types G--K according to their 2MASS and UKIDSS colors) were forced to have null \metcol colors. The typical dispersion associated with this photometric calibration method was $\pm$0.04\,mag (this represents the scatter of the $H$\,=\,13--16\,mag sources around zero methane color), which was added quadratically to the uncertainties of the targets' instrumental \metcol indices. 

Final $H$ and \metcol magnitudes are provided in Table~\ref{phottab}, where the top panel includes our photometry of the S\,Ori targets and the bottom panel contains the data of the 15 field T dwarfs in our sample. For those few targets lacking LIRIS $H$-band data, their photometry was gathered from the literature and converted into the appropriate photometric system, as it is indicated in Table~\ref{phottab}. The LIRIS $H$-band magnitudes of the field T dwarfs agree with those of the literature within 2-$\sigma$ of the quoted photometric errors. Particularly, the methane color of ULAS\,J085910.69$+$101017.1 listed in Table~\ref{phottab} deviates by only 0.04 mag with respect to the one published in \citet{pinfield08} using the same LIRIS dataset as ours. 

\begin{table}
\caption{$H$-band (MKO system) and $H$-band methane photometry}. \label{phottab}
\centering
\scriptsize
\tabcolsep=0.09cm
\begin{tabular}{llr}
\hline\hline
\noalign{\smallskip}
Object & \multicolumn{1}{c}{$H$}   & \multicolumn{1}{c}{$H-CH_{\rm 4on}$} \\
       & \multicolumn{1}{c}{(mag)} & \multicolumn{1}{c}{(mag)}           \\
\hline
\noalign{\smallskip}
S\,Ori\,70                       & 20.07\,$\pm$\,0.08 & $-$0.60\,$\pm$\,0.16   \\
S\,Ori\,73                       & 20.58\,$\pm$\,0.05\tablefootmark{a} & $-$0.11\,$\pm$\,0.15 \\
S\,Ori\,J053804.65$-$021352.5    & 19.24\,$\pm$\,0.19\tablefootmark{b} & $-$0.28\,$\pm$\,0.21 \\
\hline
\noalign{\smallskip}
2MASS\,J072718.24$+$171001.2     & 15.65\,$\pm$\,0.06   & $-$0.71\,$\pm$\,0.07   \\
2MASS\,J075547.95$+$221216.9     & 15.71\,$\pm$\,0.04   & $-$0.16\,$\pm$\,0.07   \\
2MASS\,J075840.37$+$324724.5     & 14.21\,$\pm$\,0.04   &    0.04\,$\pm$\,0.07   \\
2MASS\,J104753.85$+$212423.4     & 15.75\,$\pm$\,0.02   & $-$0.53\,$\pm$\,0.04   \\
2MASS\,J110611.97$+$275422.5     & 14.23\,$\pm$\,0.05   &    0.05\,$\pm$\,0.07   \\  
2MASS\,J120747.17$+$024424.9     & 14.56\,$\pm$\,0.10   &    0.32\,$\pm$\,0.11   \\
2MASS\,J152039.74$+$354621.0     & 14.58\,$\pm$\,0.05\tablefootmark{c} & 0.22\,$\pm$\,0.15   \\
2MASS\,J154614.61$+$493211.4     & 15.35\,$\pm$\,0.12   &    0.02\,$\pm$\,0.13   \\
2MASS\,J000013.54$+$255418.0     & 14.79\,$\pm$\,0.05   &    0.01\,$\pm$\,0.05   \\ 
2MASS\,J003451.57$+$052305.0     & 15.56\,$\pm$\,0.03   & $-$0.56\,$\pm$\,0.06   \\
ULAS\,J081948.10$+$073323.2      & 18.55\,$\pm$\,0.05   & $-$0.31\,$\pm$\,0.08   \\
ULAS\,J085139.00$+$005341.0      & 18.94\,$\pm$\,0.04   & $-$0.03\,$\pm$\,0.05   \\
ULAS\,J085342.94$+$000651.8      & 19.21\,$\pm$\,0.09   &    0.99\,$\pm$\,0.13   \\
ULAS\,J085910.69$+$101017.1      & 18.58\,$\pm$\,0.06   & $-$1.01\,$\pm$\,0.18   \\
ULAS\,J094516.40$+$075545.6      & 17.90\,$\pm$\,0.04   & $-$0.20\,$\pm$\,0.05   \\
\hline
\end{tabular}
\tablefoot{ 
\tablefoottext{a}{Taken from \citet{penaramirez11}.}
\tablefoottext{b}{Taken from \citet{penaramirez12}.}
\tablefoottext{c}{Taken from the 2MASS catalog and converted into the Manua Kea Observatory (MKO) system.}
}
\end{table}

\begin{figure}[!ht]
\centering
\includegraphics[width=0.42\textwidth]{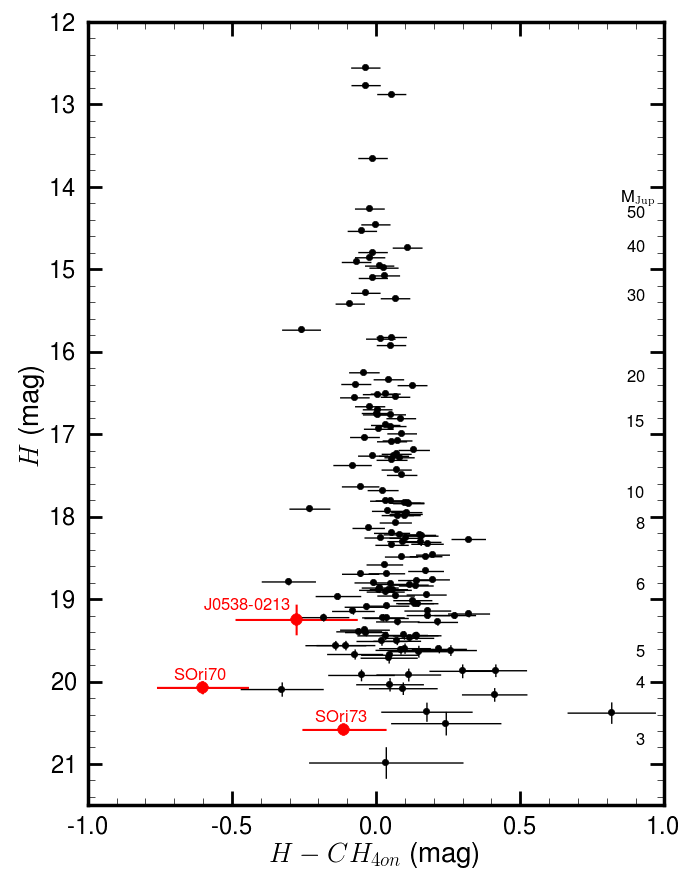}
\caption{$H$ versus $H-CH_{\rm 4on}$ color--magnitude diagram corresponding to the \cero LIRIS field. The T-type S\,Ori objects are plotted as red circles. The presence of methane absorption at 1.6\,$\mu$m induces blue LIRIS $H-CH_{\rm 4on}$ colors. Masses in units of Jupiter mass predicted for an age of 3\,Myr and the cluster distance of 352\,pc by \citet{chabrier00} are labeled.}
\label{cmplot}
\end{figure}

The $H$-band photometry of \cero agrees with the recent magnitudes reported in the literature at the 1-$\sigma$ level \citep{zapatero08}, which indicates that this object may have little photometric variability, typically below 0.1 mag in the $H$ band. \master remained undetected in our LIRIS $H$-band data. According to \citet{penaramirez12}, this source is about half a magnitude fainter than the limiting magnitude of the LIRIS image ($H_{\rm lim}$\,=\,18.7\,mag). To compute the \metcol color we used the datum in \citet{penaramirez12}. Figure~\ref{cmplot} shows the resulting $H$ versus $H-CH_{\rm 4on}$ color--magnitude diagram of S\,Ori\,70 LIRIS field. For completeness, we also added \tres and \master in the diagram. As expected, all three S\,Ori objects have blue methane colors in contrast to the nearly zero index displayed by the great majority of sources of similar brightness, which suggests the presence of methane absorption at 1.6\,$\mu$m. This is the first evidence of methane absorption in the atmosphere of J0538$-$0213. Based on the \metcol indices, $H$-band methane absorption is stronger in \cero and weaker in S\,Ori\,73. As is discussed in Section~\ref{anal_spec}, this agrees with \tres having a spectral type earlier than that of \master and S\,Ori\,70.

In this work, and for astrometric purposes (see Section~\ref{astrometry}), we also retrieved public HST/WFC3 images of S\,Ori\,70 and 73 obtained with the $F140W$ broadband filter on 2010 Sep 5 and 2010 Oct 6, respectively. This instrument has a Teledyne 1000 $\times$ 1000\,pix HgCdTe detector with a plate scale of 0\farcs128 projected onto the sky. In its imaging mode, WFC3 has a field of view of 123 $\times$ 136\,arcsec$^2$. The wide WFC3 F140W filter covers the gap between the $J$ and $H$ bands that is inaccessible from the ground. Standard calibrations were applied (dark current subtraction, linearity correction, and flat fielding) to all of the individual readouts of the WFC3 near infrared exposures. From the archive, we downloaded the final HST/WFC3 combined images that included all individual processed readouts. We also used the ISAAC $J$-band acquisition image of \master to measure this object's proper motion. It was taken immediately before the spectra described in Section~\ref{spectroscopy} on 2012 Dec 4. The pixel size was 0\farcs148, the field of view was 152\,$\times$\,152\,arcsec$^2$, and the exposure time was 60\,s. Two ISAAC images were taken and conveniently shifted for a proper subtraction of the background emission.

\section{Analysis and results}\label{anal}
\subsection{Spectroscopic classification}\label{anal_spec}
\begin{figure}[ht!]
\centering
\includegraphics[scale=0.33]{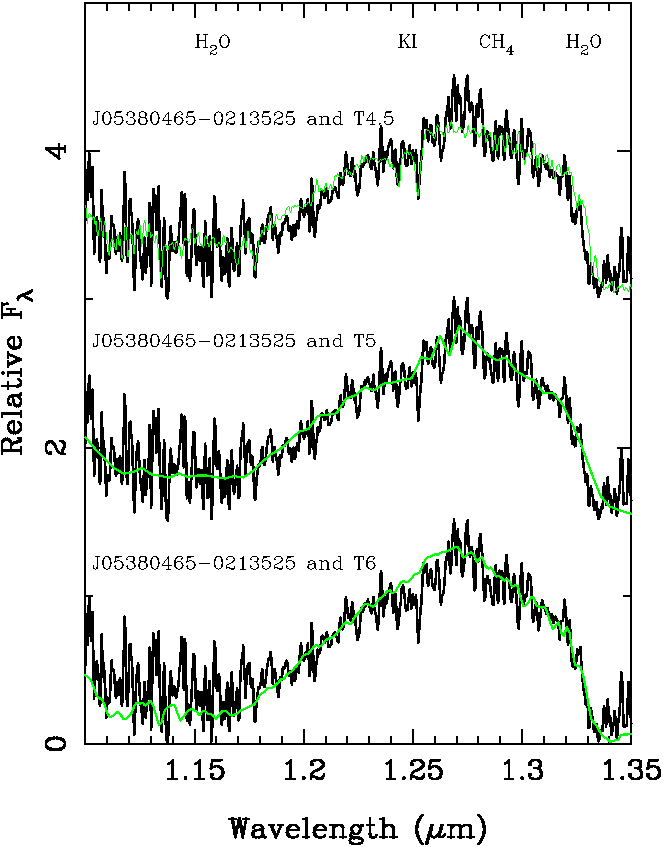}
\caption{The ISAAC $J$-band spectrum of J0538$-$0213 (black) is shown in comparison with field T dwarfs from the literature (green): 2MASS\,J05591914$-$1404488 (T4.5, \citealt{cushing05}), SDSS\,J212413.89$+$010000.3 (T5, \citealt{chiu06}), and 2MASS\,J09373487$+$2931409 (T6, \citealt{knapp04}). The T4.5 field dwarf has a similar spectral resolution to our data, while the other two field dwarfs have lower spectral resolution. The most prominent features are indicated at the top. We note that the K\,{\sc i} doublet at around 1.25\,$\mu$m is detected in J0538$-$0213, whose spectrum is shown with a smoothing of 3 pixels. All spectra are normalized to unity at 1.28--1.32\,$\mu$m. A constant offset of 1.5 and 3.0 was added for clarity.}
\label{nir}
\end{figure}

Figure~\ref{nir} illustrates the final ISAAC spectrum of J0538$-$0213. It is shown together with known field T4--T6 dwarfs from the literature. The overall spectral shape of the T-type candidate $J$-band spectrum is well matched by the field dwarfs, including the characteristic molecular absorptions due to water vapor at 1.15 and 1.33\,$\mu$m and methane at $\sim$1.28\,$\mu$m. The potassium doublet at around 1.25\,$\mu$m is detected with pseudo-equivalent widths\footnote{Equivalent width measured with respect to a pseudo-continuum.} of 3.4\,$\pm$\,1~\AA~for the K\,{\sc i} line at 1.243\,$\mu$m, and 5.2\,$\pm$\,1~\AA~for the line at 1.253\,$\mu$m. These values are consistent within the quoted error bars with those available in the literature for field T4--T6 dwarfs (see Figure 5 of \citealt{faherty14}). From the comparison shown in Figure~\ref{nir}, we derived a spectral type of T5\,$\pm$\,0.5 for J0538$-$0213, which is mainly based on the strong water vapor and methane absorption displayed in the $J$ band. The last column of Table~\ref{astrom} indicates our final spectroscopic assignments for the S\,Ori objects.

\begin{figure*}[ht!]
\centering
\includegraphics[width=0.52\textwidth]{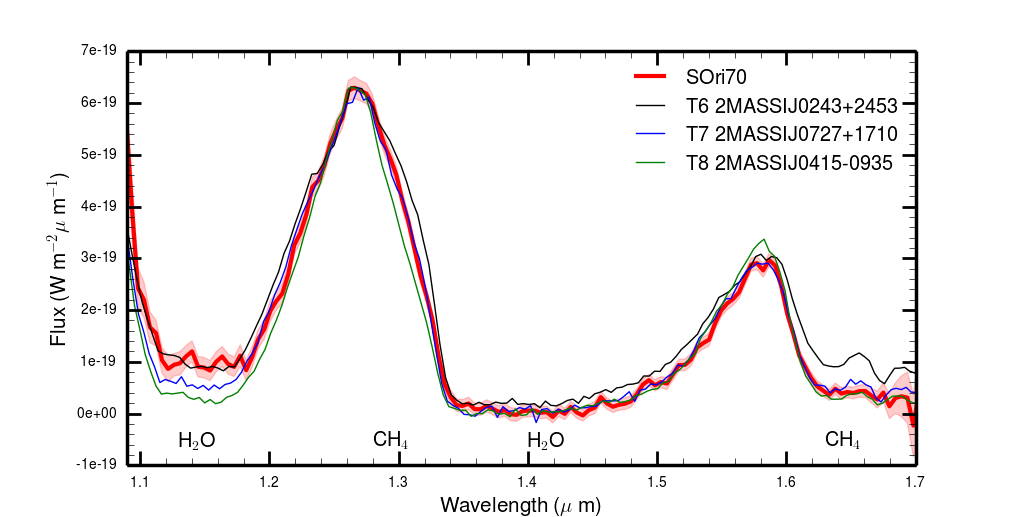}\hspace{-0.9cm}
\includegraphics[width=0.52\textwidth]{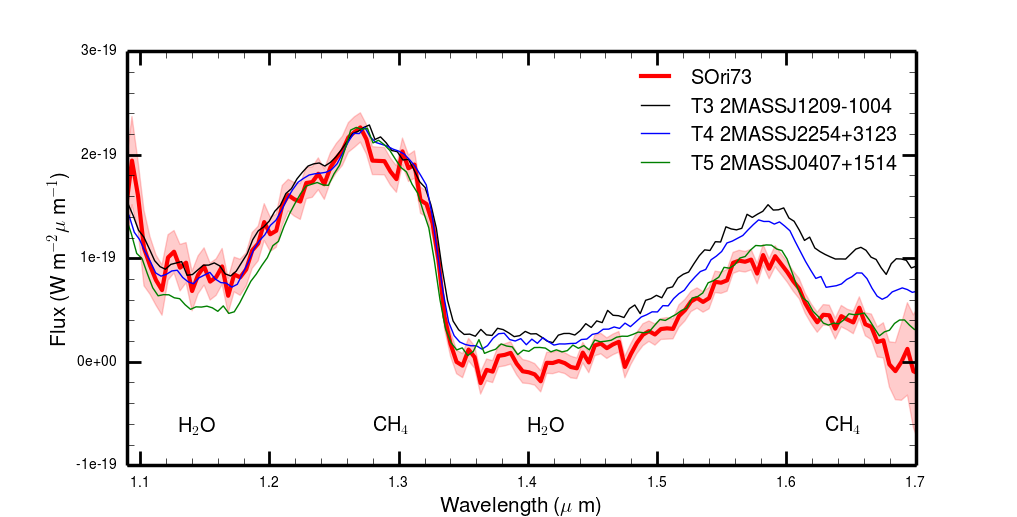}
\caption{\cero and \tres HST/WFC3 spectra (solid red line) plotted along with flux uncertainties (reddish shaded area) as given by the reduction software. Low-resolution spectra of field T dwarfs are shown with thin solid lines. In the {\sl left panel}, comparison spectra correspond to 2MASS\,IJ0243137$+$245329 (T6, \citealt{burgasser04b}), 2MASS\,IJ0727182$+$171001 (T7, \citealt{burgasser06a}), and 2MASS\,IJ0415195$-$093506 (T8, \citealt{burgasser04b}). In the {\sl right panel}, there are 2MASSJ\,12095613$-$1004008 (T3, \citealt{burgasser04b}), 2MASS\,J2254188$+$312349 (T4, \citealt{burgasser04b}), and 2MASS\,J04070885$+$1514565 (T5, \citealt{burgasser04b}). The most prominent molecular features are indicated. All spectra are normalized to the fluxes of \cero and \tres at 1.262--1.274\,$\mu$m.}
\label{hst}
\end{figure*}

The HST/WFC3 $JH$ spectra of \cero and \tres are shown in Figure~\ref{hst}. Both have features typical of the T-types, particularly the peaked-shape $H$ band, which is modeled by strong water vapor and methane absorption at the blue and red wavelengths of the band, respectively. This confirms the previous methane imaging results of \tres reported by \citet{penaramirez11}. In Figure~\ref{hst}, we also included the low-resolution spectra of various field dwarfs taken from the literature. It becomes apparent that there are no major differences between the overall shape of \cero and \tres spectra and the data of the field T dwarfs. The visual inspection of the spectra yields that the $J$ and $H$ bands of \tres are more accurately reproduced by spectral types T4 and T5. We thus assigned a typing of T4.5\,$\pm$\,0.5 to the HST/WFC3 spectrum of S\,Ori\,73. As for S\,Ori\,70, both the $J$- and $H$-band parts of the HST/WFC3 spectrum better resembles T7 classification. This determination, although slightly cooler, is consistent with the previous assignments by \citet{zapatero02sori70} and Burgasser et al$.$ (2004). \cero is the coolest object in our sample of S\,Ori objects.

\begin{figure*}[ht!]
\centering
\includegraphics[width=0.9\textwidth]{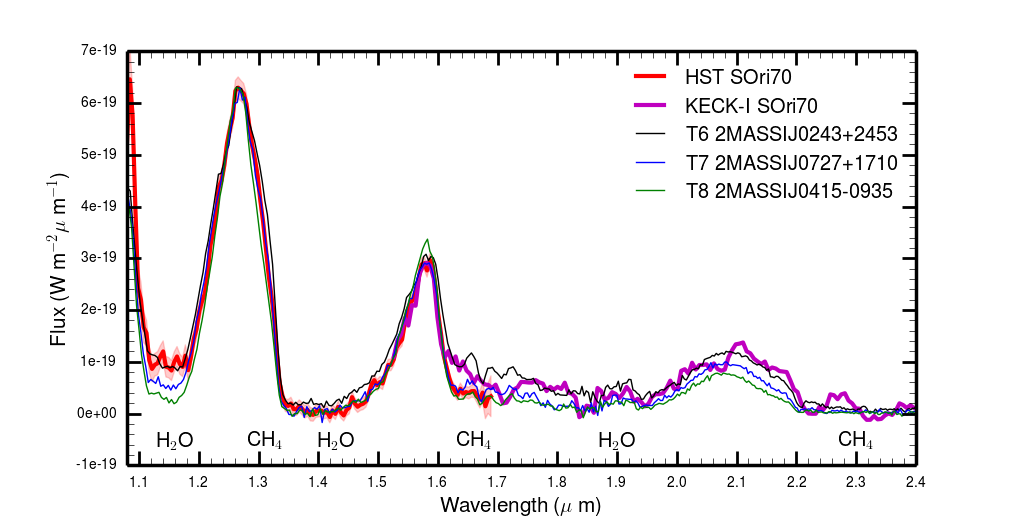}
\caption{Merged HST/WFC3 ($JH$, red) and Keck/NIRC ($HK$, magenta) spectrum of S\,Ori\,70. The two original datasets were fused at the $H$ band (1.575--1.581\,$\mu$m). Comparison spectra (thin color lines) correspond to 2MASS\,IJ0243137$+$245329 (T6, \citealt{burgasser04b}), 2MASS\,IJ0727182$+$171001 (T7, \citealt{burgasser06a}), and 2MASS\,IJ0415195$-$093506 (T8, \citealt{burgasser04b}). The most prominent features are indicated. Comparison spectra are normalized to the flux of \cero at 1.263--1.269\,$\mu$m. 
}
\label{merged}
\end{figure*}

In order to complete the near-infrared spectral coverage of \cero from the $J$ through the $K$ wavelengths (1.08--2.4\,$\mu$m), we merged the HST/WFC3 spectrum with the Keck data of \citet{zapatero02sori70}. The Keck data were obtained with a similar spectral resolution ($R \sim 90$) using the Near Infrared Camera spectrograph (NIRC, \citealt{nirc}) mounted on the Keck\,I telescope on 2001 December. The two spectra were fused in the interval 1.575--1.581\,$\mu$m within the $H$ band. The merged spectrum is illustrated in Figure~\ref{merged}. After comparison with field dwarfs, we determined that whereas the $JH$ part is nearly coincident with a T7 classification, the $K$ band resembles an earlier type (T5--T6). This agrees with the red $J-K$ color of S\,Ori\,70. In the last column of Table~\ref{astrom}, we listed T7\,$^{+0.5}_{-1.0}$ for S\,Ori\,70, i.e., that its near-infrared spectrum resembles T7 with an over-luminosity in the $K$ band.

\begin{table}
\caption{Proper motion and spectral types of S\,Ori objects. \label{astrom}}
\centering
\scriptsize
\tabcolsep=0.09cm
\begin{tabular}{lccccc}
\hline\hline
\noalign{\smallskip}
Object & $\Delta (t)$ & $\mu_\alpha \cos\delta$  & $\mu_\delta$ & SpT\tablefootmark{a} & SpT\tablefootmark{b} \\
 & (yr) & (mas\,yr$^{-1}$) & (mas\,yr$^{-1}$) & (color) & (spec)  \\
\hline
\noalign{\smallskip}
S\,Ori\,70                     & 8.68 &   19.8\,$\pm$\,5.9 &   33.7\,$\pm$\,6.2  &  T6.6--T7.2 & T7.0\,$^{+0.5}_{-1.0}$ \\
S\,Ori\,73                     & 8.82 &   43.6\,$\pm$\,6.7 & $-$3.6\,$\pm$\,7.7   &  T3.2--T5.4 & T4.5\,$\pm$\,0.5     \\ 
S\,Ori\,J053804.65$-$021352.5  & 7.03 & $-$1.2\,$\pm$\,6.1 & $-$4.8\,$\pm$\,7.5   &  T4.5--T6.5 & T5.0\,$\pm$\,0.5   \\
\hline
\end{tabular}
\tablefoot{ 
\tablefoottext{a}{This paper, based on LIRIS methane colors.}
\tablefoottext{b}{This paper, based on low-resolution spectra.}
}
\end{table}

\begin{table}
\caption{Average LIRIS methane colors for different spectral types. \label{photFL}}
\centering
\scriptsize
\tabcolsep=0.09cm
\begin{tabular}{lcr}
\hline\hline
\noalign{\smallskip}
SpT & \metcol  &N\tablefootmark{a}\\
 &(mag)&\\
\hline
\noalign{\smallskip}
F0--F5           & 0.001\,$\pm$\,0.007    & 7  \\
F5--F9.5         & 0.003\,$\pm$\,0.006    & 7  \\
G0--G5           & 0.018\,$\pm$\,0.010    & 7  \\
G5--G8           & 0.019\,$\pm$\,0.009    & 4  \\
K0--K5           & 0.044\,$\pm$\,0.009    & 6  \\
K5--K8           & 0.057\,$\pm$\,0.011    & 4  \\
M0--M5           & 0.091\,$\pm$\,0.013    & 14 \\
M5--M9           & 0.153\,$\pm$\,0.025    & 11 \\
L0--L5           & 0.207\,$\pm$\,0.017    & 8  \\
L5--L9           & 0.209\,$\pm$\,0.007    & 6  \\
\hline
\end{tabular}
\tablefoot{ 
\tablefoottext{a}{Number of sources used to compute the average color.}
}
\end{table}

\subsection{Methane filter spectral types}\label{anal_phot}
We also derived the spectral types of the S\,Ori objects using the methane colors in Table~\ref{phottab}. Prior to this derivation, the LIRIS \metcol colors must be calibrated against spectral type. For this purpose, we followed two approaches. One was the photometric measurement of \metcol indices of a sample of 15 field T dwarfs (Section~\ref{targets}) using data obtained with an identical instrumental configuration to that of our main targets. This photometry is given in Table~\ref{phottab}. The other method consisted in the integration of tens of observed spectra of field dwarfs convolved with the LIRIS filters response curves. Both approaches yielded consistent results and are jointly described below. 

\begin{figure}[!ht]
\centering
\includegraphics[width=0.44\textwidth]{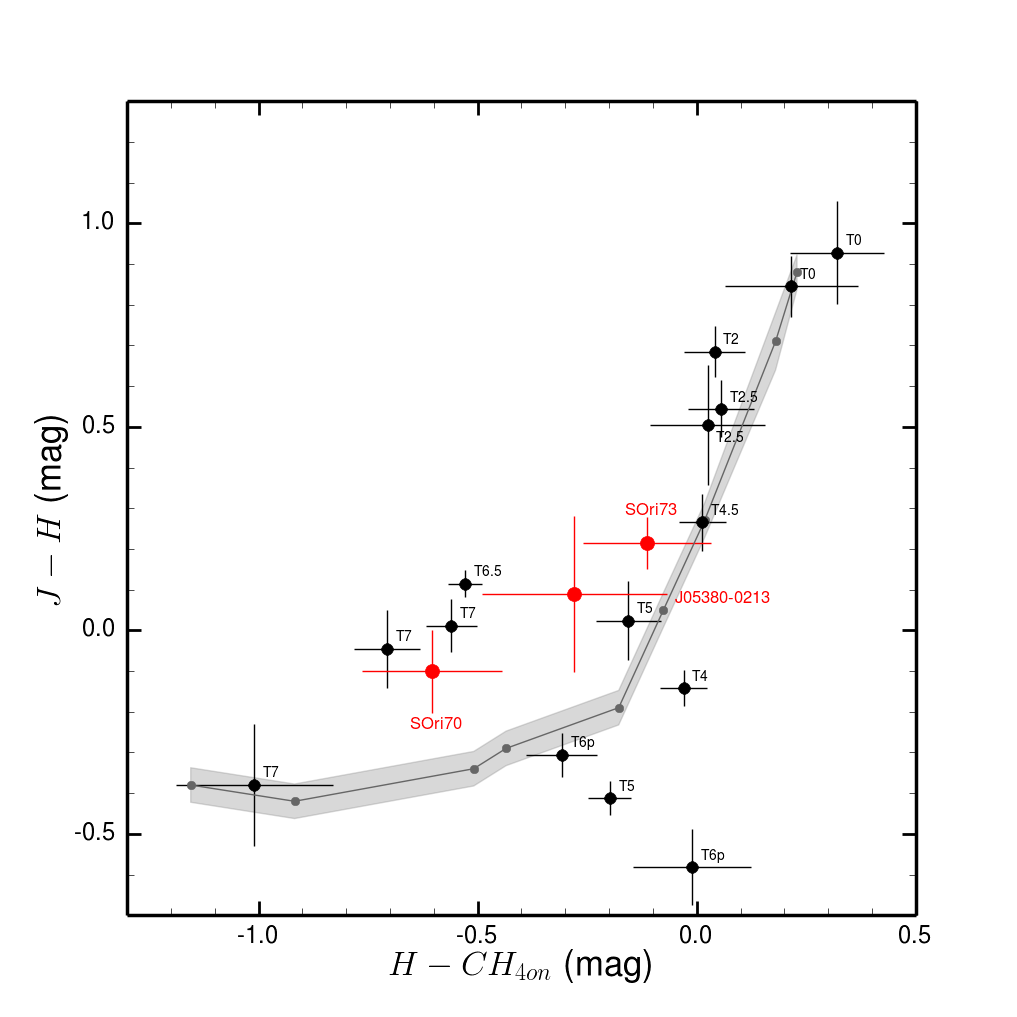} \\
\includegraphics[width=0.44\textwidth]{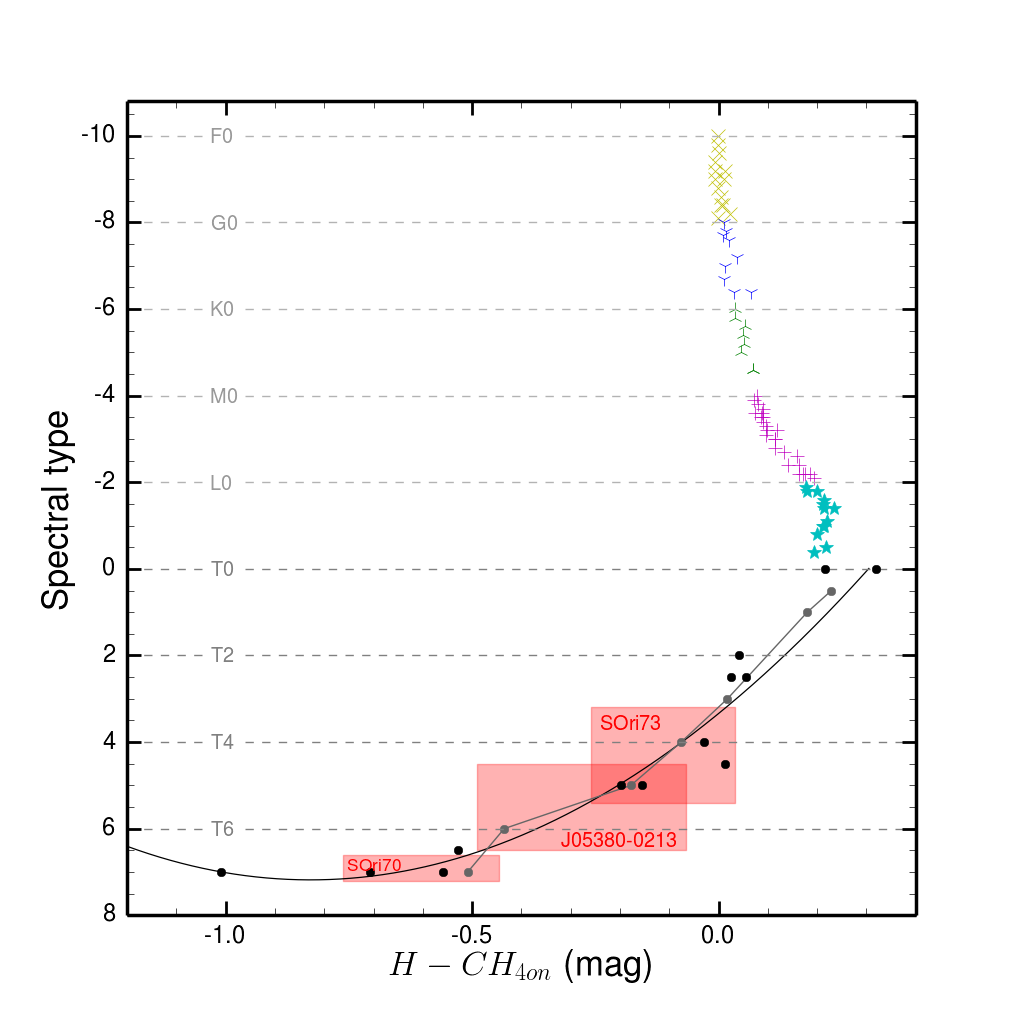}
\caption{The \metcol color is shown as a function of $J-H$ of T-type dwarfs in the {\sl top panel}, and as a function of F0 through T7 spectral types in the {\sl bottom panel}. S\,Ori\,70, 73, and J0538$-$0213 are labeled and indicated in red. The black solid circles stand for the 15 field T dwarfs with LIRIS photometric data (the T6p dwarf is not included in the bottom panel). In the {\sl top panel}, the solid line with a gray shaded area (accounting for $J-H$ errors only) corresponds to the spectrophotometric methane colors computed from the integration of literature observed spectra (from right to left, the small dots in this line represent T0.5, T1, T3, T4, T5, T6, T7, T8, and T9). We note the agreement between this curve and the directly measured LIRIS colors. This line is also included in the {\sl bottom panel} in gray and overlaps with the solid black curve that represents the second-order polynomial fit of the LIRIS methane observations. The $y$-axis is not equally scaled for all spectral types. For the T-type regime, the scale is enlarged for the clarity of the figure. The reddish boxes of the S\,Ori objects stand for the LIRIS \metcol error bars and the uncertainties associated with the determination of their spectral types from the methane data. }
\label{main}
\end{figure}

We retrieved observed, good quality spectra of field objects classified as F- through T-type dwarfs from the catalogs of \citet{rayner09}, \citet{cushing05}, \citet{knapp04}, \citet{chiu06}, and \citet{golimowski04}. All retrieved spectra were conveniently convolved with the LIRIS $H$ and \met filter transmission profiles and were flux calibrated to derive the \metcol indices. Table~\ref{photFL} provides the resulting spectrophotometric \metcol colors for different intervals of spectral types; the indices have associated errors computed as the standard deviation of the spectrophotometric measurements. The last column of Table~\ref{photFL} gives the number of total objects used per spectral type interval. Figure~\ref{main} depicts the spectrophotometric \metcol indices as a function of $J-H$ colors (top panel) and spectral type (bottom panel). The top panel focuses on the T-types. Whereas the F- and G-type stars display a zero methane color within the quoted error bars, \metcol slowly becomes redder for later spectral types. The reddest indices correspond to the late L types (bottom panel of Figure~\ref{main}). Early T dwarfs also have red methane colors, which rapidly turn over into blue values by mid-T as a consequence of the atmospheric methane absorption in the $H$ band. LIRIS methane imaging can be used to easily identify T dwarfs with spectral types $\sim$T3 and later. 

The LIRIS photometric magnitudes of the S\,Ori objects and the sample of 15 T dwarfs are also shown in both panels of Figure~\ref{main}. The positions of S\,Ori\,70, 73, and \master in the top panel are consistent with spectral types T6--T7, T4--T5, and T4--T7, respectively. The error bar associated with \master is the largest because of its high uncertainty in the \metcol color. For future references, we applied a second-order polynomial fit to the \metcol indices of the field T dwarfs observed with LIRIS and obtained the following equation that relates T spectral type and the $H$-band methane colors,
\begin{equation}
{\rm SpT}  = 3.34 - 9.28  (H-CH_{\rm 4on}) - 5.602  (H-CH_{\rm 4on})^2,
\label{eq}
\end{equation}
where T0, T1, ... corresponds to SpT = 0, 1, ... This equation is valid for spectral types in the interval T0--T7, has a root mean square (rms) of 0.29 subtypes, and is plotted as a solid thick line in the bottom panel of Figure~\ref{main}. According to Eq.~\ref{eq}, we determined the following spectral types based on the observed methane colors and their errors: T6.6--T7.2 for S\,Ori\,70, T3.2--T5.4 for S\,Ori\,73, and T4.5--T6.5 for J0538$-$0213. These values are listed in Table~\ref{astrom}. They are fully consistent with the spectroscopic determinations at the 1-$\sigma$ level. \master is confirmed as having an intermediate spectral type between \tres and S\,Ori\,70. 

\subsection{Astrometry}\label{astrometry}
To obtain proper motions, we used the $K_s$-band Keck/NIRC image of \cero obtained by \citep{zapatero02sori70} on 2002 Feb 25 as the first epoch data, and the HST/WFC3 F140W image described in Section~\ref{methane_imaging} as the second epoch data. The Keck/NIRC image had a pixel size of 0\farcs128. \cero was detected with a signal-to-noise ratio (S/N) of 7 and 286 in both images (S/N was measured as the ratio between the peak of the object's radial profile and the noise of the subtracted background contribution). The ISAAC $J$-band image (2001 Dec 10, pixel scale 0\farcs15) of \citet{caballero07} and the HST/WFC3 F140W image of \tres acted as the first and second epoch data. \tres was detected with S/N = 11 and 166, respectively. As for J0538$-$0213, the first epoch image (2005 Nov 22, pixel scale of 0\farcs4) was taken from the UKIDSS archive ($J$-band image) and the second epoch frame was acquired with ISAAC (see Section~\ref{methane_imaging}). This source was detected with S/N = 5 and 14, respectively. For all three targets, the second column of Table~\ref{astrom} provides the time intervals between the two epochs of observations; they lie in the range 7--9\,yr. The seeing of all data employed for the proper motion analysis is typically $\le$1\arcsec.

Relative proper motions were measured by comparing the relative positions of the targets with respect to 6--20 unresolved reference sources within an area of 0.25--6.25\,arcmin$^2$ (depending on the fields of view of the different instruments). All reference sources were detected with S/N higher than 15 in both epochs. Pixel coordinates were transformed using third-order polynomial fits (except for S\,Ori\,70, for which a second-order polynomial was employed) and the {\sc geomap} routines within {\sc iraf}. These fits were expected to correct for field distortions and rotation angles, and different plate scales. Coordinates transformations were achieved with a typical precision of $\pm$0.2 pix for the $x$- and $y$-axes (a 3-$\sigma$ rejection algorithm was used, where $\sigma$ denotes the dispersion of the transformations). Our targets were not included in the definition of the transformation equations for null motion. The resulting pixel shifts of the S\,Ori objects were converted into proper motions ($\mu_\alpha \cos\delta$, $\mu_\delta$) by taking into account the two-epochs time differences, pixel scales, and the orientations of the frames. 

To convert our relative proper motions into absolute values, we estimated the
required corrections from the expected average motions of a sample of 20 stars in
the magnitude ($J \sim 14-17$ mag) and coordinate intervals of the astrometric
reference sources. This sample was simulated using the Besan\c con model of
stellar population synthesis of the Galaxy \citep{besancon}. We found average
proper motions of $\mu_\alpha\,{\rm cos}\,\delta$ = $-0.2 \pm 3.0$ and
$\mu_\delta$ = $-2.9 \pm 4.3$  mas\,yr$^{-1}$, where the error bars correspond to
the dispersion of the predicted motions. These values agree with the mean motion of cluster non--member stars in 
the field of $\sigma$\,Orionis published by \citet{dias06,dias14}, and are consistent with a null velocity as expected for
randomly distributed small proper motions. Therefore, we did not apply any
correction and our final astrometric measurements of the S\,Ori targets are given
in Table~3; the relative proper motion uncertainties were calculated by
quadratically adding the dispersion of the polynomial transformations and the
errors of the targets centroids (which were $\pm$0.02--0.13 pix). To account for
the uncertainty associated with the relative to absolute proper motion conversion,
we linearly added the dispersion of the mean proper motions of the astrometric
reference stars as given by the Besan\c con model to the final astrometric errors
shown in Table~3. The motions of S\,Ori\,70 and 73 fully agree with the values of
\citet{penaramirez11} at the level of 0.3-$\sigma$ and 0.2-$\sigma$,
respectively. We note that in \citet{penaramirez11} we provided the
uncertainties associated with the relative proper motions while here we report the
error bars of the absolute measurements.

\begin{figure}[ht!]
\centering
\includegraphics[width=0.5\textwidth]{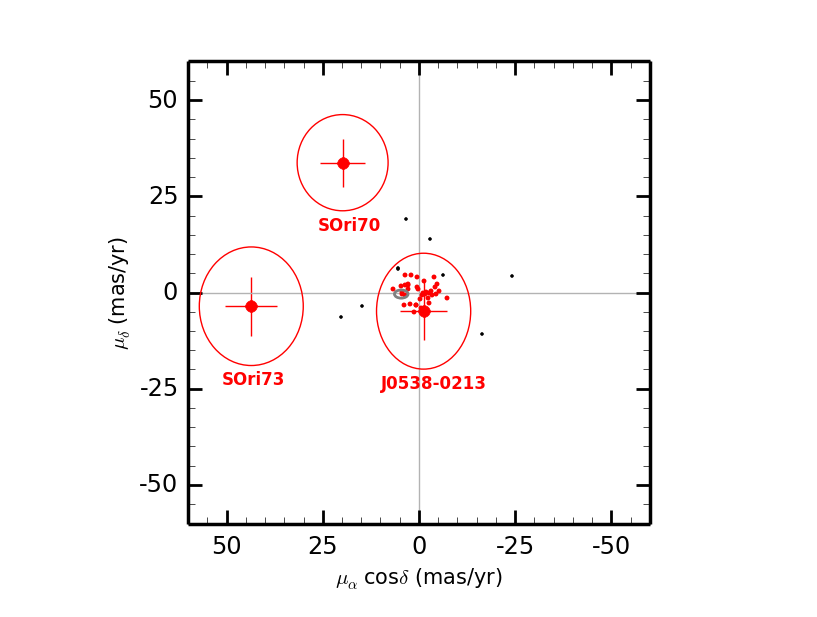}
\caption{Proper motion diagram of S\,Ori\,70, S\,Ori\,73, and J0538$-$0213 (solid circles and labeled). All the reference sources within an area of 0.25\,arcmin$^2$ (S\,Ori\,70), 4.60\,arcmin$^2$ (S\,Ori\,73), and 6.25\,arcmin$^2$ (J0538$-$0213) are plotted as small dots; those defining the astrometric transformation equations are in red. The ellipses around the S\,Ori targets represent their 2-$\sigma$ proper motion uncertainties. The \textit{Hipparcos} motion of the \so cluster is depicted with a gray solid-line ellipse.}
\label{pms}
\end{figure}

Figure~\ref{pms} illustrates the resulting proper motion diagram for S\,Ori\,70, 73, and J0538$-$0213. The three objects have clearly distinct motions: $\mu$\,=\,39.1\,$\pm$\,8.6\,mas\,yr$^{-1}$ and position angle ($PA$, measured east of north) of 30\fdg5\,$\pm$\,2\fdg3 (S\,Ori\,70), $\mu$\,=\,43.7\,$\pm$\,10.2\,mas\,yr$^{-1}$ and $PA$ = 94\fdg7\,$\pm$\,9\fdg6 (S\,Ori\,73), and $\mu$\,=\,5.0\,$\pm$\,9.7\,mas\,yr$^{-1}$ and $PA$ = 194\fdg0\,$\pm$\,42\fdg6 (J0538$-$0213).

To assess the astrometric membership of the studied T-type sources in the
$\sigma$\,Orionis cluster, we adopted the {\sl Hipparcos} proper motion of the
multiple star $\sigma$\,Ori \citep{perryman97} because it is consistent within
error bars with other independent determinations by different groups
\citep{kharchenko05,dias06,dias14,caballero07a} and has a small error bar. Only
\citet{vanleeuwen07} provided a differing proper motion. We confirmed that \cero and 73 have motions larger than the $\sigma$~Orionis cluster ($\mu$\,=\,4.7\,$\pm$\,1.0\,mas\,yr$^{-1}$, $PA$ = 95\fdg0\,$\pm$\,10\fdg0, \citealt{perryman97}) by $>$3.5-$\sigma$ (the first epoch images employed here were the same ones used by \citealt{penaramirez11}). We note that this is the first proper motion measurement of J0538$-$0213; this T5 dwarf does not appear to have a significant motion since its proper motion is consistent with a null displacement in the time interval of the observations. 

\section{Discussion and final remarks}\label{discussion}
All three $\sigma$~Orionis T-type candidate members are confirmed as having significant methane absorption in their atmospheres consistent with spectral types ranging from T4.5 to T7. However, 
there is little support for their cluster membership from the currently available photometric, astrometric, and spectroscopic data. \master displays the K\,{\sc i} doublet at around 1.25\,$\mu$m with a strength comparable to that of field T4--T6 dwarfs. In addition, its near- and mid-infrared photometry and $H$-band methane color are consistent with the spectral type measured from the $J$ wavelengths and with the properties observed in high-gravity field dwarfs. At the resolution of our spectra ($R \sim 50-90$), neither \cero nor \tres appears to deviate significantly from T6--T7 and $\sim$T4 field dwarfs, except for S\,Ori\,70, whose $K$-band spectrum indicates an earlier spectral classification consistent with its red near-infrared colors. This effect can be explained by the collision-induced absorption (CIA) by H$_2$ that affects this region of the spectrum. As pointed out by \citet{saumon12}, CIA H$_2$ opacity is expected to be reduced for objects with a lower gravity and/or greater metallicity. 

We caution about the interpretation of spectroscopic features due to low gravity in very low-resolution T-type spectra. First of all, the impact of low gravity on the output energy distribution of methane dwarfs is not well established observationally. If any marked deviation had been seen in the spectra of one of the S\,Ori objects, we would have immediately linked it to a low-gravity atmosphere. However, no evidence of significant spectroscopic deviations does not necessarily imply high gravities. Some of the youngest T-type dwarfs with reasonably constrained ages are the T3.5 GU\,Psc\,b \citep{naud14}, the T7 CFBDSIR\,J214947.2$-$040308.9 \citep{delorme12}, and the $\sim$T9 GJ 504b \citep{kuzuhara13}, whose ages are believed to be around 100 (AB~Doradus stellar moving group) and 160\,Myr (about 30--50 times older than $\sigma$~Orionis). According to the discovery papers and \citet{janson13}, all these objects have spectral (methane) properties similar to those of field dwarfs believed to be much older. While the colors of the T3.5 GU\,Psc\,b resemble those of field dwarfs of similar classification, the later T7 CFBDSIR\,J214947.2$-$040308.9 and $\sim$T9 GJ 504b are anomalously bright in the $K$-band wavelengths, which could be a signature of high metallicity or possibly low surface gravity. This redness of the $K$ band relative to field brown dwarfs of similar temperatures or types is a feature that is also present in other objects with indications of low gravity, such as Ross 458\,C \citep{goldman10,burningham11}. S\,Ori\,70 can be added to this list.

Regarding proper motions, only \master displays a motion that is consistent with that of $\sigma$~Orionis at the 0.9-$\sigma$ level. Nevertheless, given the position of the cluster with respect to our location in the Galaxy, the coincidence of proper motions cannot be used as a robust criterion to assess cluster membership. At the distance of 352\,pc, the tangential velocities of \cero and \tres are discrepant with the velocities of the bright $\sigma$~Orionis members, which suggests they are not members of the cluster, but rather probable foreground objects. Using our spectral types determinations and the $H$-band photometry of S\,Ori\,70, 73, and J0538$-$0213, and the absolute magnitudes of field T dwarfs given by \citet{caballero08contamination} and \citet{metchev08}, we inferred that in the scenario of high-gravity atmospheres they would be located at 54--77\,pc (S\,Ori\,70), 134--195\,pc (S\,Ori\,73), and 61--80\,pc (J0538$-$0213), in which case their tangential velocities would be close to those of field T dwarfs measured by \citet{faherty09}. Given the similarity of the updated proper motions of \cero and 73 with the values of \citet{penaramirez11}, the discussion of the backward projected motion of these sources presented by these authors is still valid, and only the \cero trajectory passed through the outskirts of a star-forming region south of $\sigma$~Orionis about 6.0\,$\times$\,10$^5$\,yr ago, where it might have originated.

The $\sigma$ Orionis mass function presented by \citet{penaramirez12} does not change with the results of this work. Neither \cero nor 73 was included in the cluster mass function derivation because they were not detected in the $Z$ band of the survey (that mass function was obtained for cluster candidate members with clear detections in at least $Z$ and $J$ bands), and \master was included in the most massive bin within the planetary-mass regime. Removing one object does not make a significant difference in that particular mass interval and is included within the quoted error bars. The survey presented by \citet{penaramirez12} covered an area of 2798.4\,arcmin$^2$. All of our three T-type targets were detected in the $J$ band down to a magnitude of $J \simeq 21$\,mag. The photometric survey was sensitive to $\sim$T7 spectral type. Considering field ages, this implies that the exploration was complete for T4--T7 dwarfs in a volume of 752.5\,pc$^3$ or up to an heliocentric distance of 213\,pc (T4) and 119\,pc (T7). If we assume that none of the three S\,Ori sources are young but  are field dwarfs lying towards Orion, we would derive an observed T4--T7 dwarf object density of 4.1\,$\pm$\,2.7\,$\times$\,10$^{-3}$\,pc$^{-3}$ (or 2.8\,$\pm$\,1.6\,$\times$\,10$^{-3}$\,pc$^{-3}$ if at least one is young). This object's density is similar, although slightly larger by a factor of 1.1--2.5, to the volume densities published by \citet{metchev08}, \citet{caballero08contamination}, and \citet{reyle10}. 

Recently, \citet{zapatero14} has reported that the L-T transition objects in the intermediate-age Pleiades cluster (120\,Myr) appear at magnitudes and colors fainter and redder than expected. This is particularly interesting in the context that thick clouds may form at low gravity and low mass, which thus affect the brightness of young objects by increasing the cloud opacity and making young lower-mass objects fainter at optical and near-infrared wavelengths. This scenario is explained by \citet{saumon08} and \citet{marley12}. By considering how much fainter the Pleiades L-T dwarfs are in the $J$ band, this would locate the $\sigma$~Orionis L-T transition and the cluster methane dwarfs at $J \ge 21$ mag, which is close to the detection limit of the \citet{penaramirez12} survey. Following the discussion of \citet{penaramirez12} and the likely non-membership of the three T-type candidates targeted in this work, we can conclude that either the $\sigma$~Orionis mass function has a mass cut-off at around the planetary mass of $\sim$4 M$_{\rm Jup}$, or the cluster T dwarfs lie at magnitudes fainter than expected and the cluster mass cut-off for the formation of free-floating planets is yet to be determined. Deeper explorations of $\sigma$~Orionis or similarly young clusters or star-forming regions are needed to solve this ambiguity.

\begin{acknowledgements}
We thank Philip Lucas for agreeing to let us use the HST spectra. We also
thank him for refereeing this manuscript and providing useful comments. This work is based on observations made with the WHT, installed at the Spanish Observatorio del Roque de los Muchachos in the island of La Palma, Spain, and is also based on observations made with ESO Telescopes at the Paranal Observatory under program ID 090.C-0766, and on observations made with the NASA/ESA Hubble Space Telescope, obtained from the data archive at the Space Telescope Science Institute. STScI is operated by the Association of Universities for Research in Astronomy, Inc. under NASA contract NAS 5-26555. This work is partly funded by the Spanish Ministry of Economy and Competitivity through the projects AYA2011-30147-C03-03, AYA2010-20535, and the Chilean FONDECYT Postdoctoral grant 3140351. 
\end{acknowledgements}

\bibliographystyle{aa} 
\bibliography{methane_2014_printer}

\end{document}